\documentclass[a4paper,USenglish]{lipics-v2021}
\usepackage[utf8]{inputenc}
\usepackage{placeins}
\def\lf{\tiny}
\newcounter{mylinenumber}

\def\nnll{\refstepcounter{mylinenumber}\lf\themylinenumber}

\newcommand{\ignore}[1]{}

\usepackage{makecell}

\title{\textsf{RandSolomon}: Optimally Resilient Random Number Generator with Deterministic Termination}
\date{\today}



\funding{Luciano Freitas de Souza was supported by Nomadic Labs, Petr Kuznetsov and Andrei Tonkikh---by TrustShare Innovation Chair.}

\author{Luciano Freitas de Souza}{CEA LIST, Université de Paris-Saclay; LTCI, T\'el\'ecom Paris, Institut Polytechnique de Paris}{lfreitas@telecom-paris.fr}{}{}
\author{Andrei Tonkikh}{LTCI, T\'el\'ecom Paris, Institut Polytechnique de Paris}{andrei.tonkikh@telecom-paris.fr}{}{}
\author{Sara Tucci-Piergiovanni}
{CEA LIST, Université de Paris-Saclay}{sara.tucci@cea.fr}{}{}
\author{Renaud Sirdey}
{CEA LIST, Université de Paris-Saclay}{renaud.sirdey@cea.fr}{}{}
\author{Oana Stan}
{CEA LIST, Université de Paris-Saclay}{oana.stan@cea.fr}{}{}
\author{Nicolas Quero}
{CEA LIST, Université de Paris-Saclay}{nicolas.quero2@cea.fr}{}{}
\author{Petr Kuznetsov}{LTCI, T\'el\'ecom Paris, Institut Polytechnique de Paris}{petr.kuznetsov@telecom-paris.fr}{}{}

\Copyright{L. Freitas de Souza, A. Tonkikh, S. Tucci-Piergiovanni, R. Sirdey, O. Stan, N. Quero, P. Kuznetsov}

\authorrunning{L. Freitas de Souza et al.}
\titlerunning{RandSolomon}

\nolinenumbers

\begin{document}

\maketitle

\begin{abstract}
    Multi-party random number generation is a key building-block in many practical protocols. While straightforward to solve when all parties are trusted to behave correctly, the problem becomes much more difficult in the presence of faults. This paper presents \textsf{RandSolomon}, a partially synchronous protocol that allows a system of $N$ processes to produce an unpredictable common random number shared by correct participants. The protocol is optimally resilient, as it allows up to $f=\lfloor \frac{N-1}{3} \rfloor$ of the processes to behave arbitrarily, ensures deterministic termination  and, contrary to prior solutions, does not, at any point, expect faulty processes to be responsive.
\end{abstract}

\section{Introduction}
\label{sec:related}
\begin{sloppypar}
In a Byzantine fault-tolerant random number generator (\textsf{BFT-RNG}) protocol, a set of participating processes agree on a single random number that cannot be manipulated 
or halted, despite the presence of Byzantine failures, i.e., assuming that a faulty process may arbitrarily deviate from the prescribed algorithm. 
We distinguish between \emph{commission} and \emph{omission} failures~\cite{haeberlen2009fault}.
Intuitively, a \emph{commission} fault occurs when a process sends messages a correct process would not send, whereas an \emph{omission} fault occurs when a process does not send messages a correct process would send.

A BFT-RNG protocol is typically divided into three phases:

\begin{enumerate}
    \item{\bf Generation and Commitment Phase} - each process locally generates some random value and then publicly commits to this value without revealing it.
    \item{\bf Reveal Phase} - the values previously committed are revealed.
    \item{\bf Computation Phase} - using the values revealed, the processes decide on the resulting random number.
    \end{enumerate}

The idea is to make sure that at the moment the committed random values are revealed, it is already too late for the adversary to manipulate the output.
Furthermore, assuming that the local random numbers are uniformly distributed, so should be the distribution of the output.

\end{sloppypar}

To the best of our knowledge, this paper describes the first partially synchronous \textsf{BFT-RNG} protocol that maintains optimal resilience (up to $\lfloor \frac{N-1}{3} \rfloor$ Byzantine processes in a system of $N$) that ensures 
\emph{deterministic} termination. 
Unlike prior solutions~\cite{cryptoeprint:2000:034,syta2017scalable}, our protocol does not expect that faulty processes remain responsive in the generation phase, i.e., it tolerates omission faults.   
\ignore{
We achieve this in a novel way by relying on deterministic encryption and deterministic codes which allows processes to retrace the steps taken by others at the appropriate moment and check correctness.
}

\subparagraph*{State of the art.}
In designing a \textsf{BFT-RNG} algorithm, we face two major chal\-len\-ges: (i)~how to share random inputs despite omission failures, so that Byzantine processes cannot learn them before the reveal phase begins,   
and (ii)~how to compute correct results despite commission faults of Byzantine processes. 
Existing protocols solve the first challenge by using techniques such as secret sharing \cite{shamir1979share}, verifiable delay functions \cite{boneh2018verifiable}, threshold signatures \cite{bellare2006multi}\cite{boneh2001short}, and fully homomorphic encryption \cite{gentry2009fully} and the second---by requiring a verifiable proof that a shared data was generated correctly.

\vspace{2mm}\noindent\emph{Techniques}. A $(f,N)$-secret sharing \cite{shamir1979share} scheme allows a process during the generation and commitment phase to share a secret $s$ with $N$ processes so that any subset of size $f+1$ among them can retrieve $s$, while no subset of $f$ or less can. 
This way, even if a process refuses to disclose the original secret it has committed, the correct processes in the system can still reconstruct it in the reveal phase by using the shares they received earlier. 
Moreover, the values cannot be learned too early as the number of shares held by the Byzantine processes does not surpass $f$. 
%
 \emph{Threshold-signature} schemes, such as Schnorr \cite{bellare2006multi} or BLS \cite{boneh2001short}, are also very helpful in this context, as they allow to efficiently verify that a number of processes surpassing a given threshold agree with a certain value.

One can also make sure that the processes commit to a value without revealing it beforehand and provide a mechanism to retrieve commitments of Byzantine processes by using 
\emph{verifiable delay functions}~\cite{boneh2018verifiable}. 
This technique guarantees that Byzantine processes cannot use the data shared by the correct processes to change change their inputs and affect the result.  
%
Once a stipulated verifiable delay has expired, the correct processes can access the information presented by any process guaranteeing that the protocol is not halted. 

The two homomorphic structures of most interest for \textsf{BFT-RNG} are Fully Homomorphic Encryption (FHE) \cite{gentry2009fully} and homomorphic hashes. Given two sets $A$ and $B$, a map $f: A \to B$ is said to be ($\circ$-)homomorphic if it preserves an existing operation $\circ$ on both sets: $\forall x,y \in A, f(x \circ y) = f(x) \circ f(y)$ \cite{bronshtein2013handbook}.  FHE allows processes to make operations in ciphertexts without knowing the plaintexts and can be then used instead of secret sharing for solving the same problem of preventing misbehaving parties from accessing data too early on and denying the access of correct participants to the data when it must be shared. As for homomorphic hashes, they are, as the name indicates, hash functions with homomorphic properties (i.e. by performing some operations over some data and their associated hashes, one obtains a result and a consistent associated hash). Homomorphic hashes allow to solve the second challenge of \textsf{BFT-RNG} design: they provide a mean to check that an operation was correctly executed by observing the hashes of the inputs and the hash of the outputs and can therefore contribute in detecting commission failures.

Other kinds of proofs of well formed data include Verifiable Random Functions (VRF) or Public Verifiable Secret Sharing (PVSS). VRF \cite{micali1999verifiable} are functions that once provided with an input $x$, output both a random number $y$ and a proof $\pi$ that allows any process using $\pi$ to verify whether $y$ was generated using $x$ or not. Algorand's VRF \cite{10.1145/3132747.3132757} uses a common coin (generated by the Algorand consensus) to correctly generate verifiable random numbers. 
PVSS-based proof \cite{schoenmakers1999simple} exchange together with secret shares some additional information that prove the data integrity without revealing any information of the original secret.

\begin{sloppypar}
\vspace{2mm}\noindent\emph{Protocols.}
In \autoref{tab:comp}, which is a modified and expanded version of the table given in~\cite{cryptoeprint:2018:319}, we present a comparison including several existing \textsf{BFT-RNG} algorithms and the solution we present in this paper: \textsf{RandSolomon}. In some of these protocols, the networks (with $N$ nodes) are partitioned into clusters of size $c$, this parameter appears in some of the complexity bounds given in the table. 
\end{sloppypar}

\begin{table}[ht]
    \begin{center}
\resizebox{\textwidth}{!}{
  \renewcommand{\arraystretch}{2}
 \begin{tabular}{c c c c c c c c}
 \hline
 RNG & Sync. & \makecell{Vulnerability} & \makecell{Term.} & \makecell{Communication \\ Complexity \\ (Overall) }& \makecell{Computation \\ Complexity \\ (per process)} & Resilience & Techniques\\ [0.5ex] 
 \hline\hline
   Cachin et al.~\cite{cryptoeprint:2000:034}& A & \makecell{Trusted \\ key dealer}& Det. & $O(N^2)$ & $O(N)$ & $f < \frac{N}{3}$ & \makecell{Unique threshold \\ signatures (eg BLS)\cite{boneh2001short}}\\ \hline
  RandShare \cite{syta2017scalable} & A & \makecell{No ommission \\ in commit.} & Det. & $O(N^3)$ & $O(N^3)$ & $f < \frac{N}{3}$ & PVSS \cite{schoenmakers1999simple}\\ \hline  
 RandHound\cite{syta2017scalable}& A & \makecell{No ommission \\ in commit.}& \makecell{Prob.} & $O(c^2N)$ & $O(c^2N)$ & $f < \frac{N}{3}$ & \makecell{PVSS \cite{schoenmakers1999simple}\\ Multisignatures \cite{bellare2006multi}}\\ \hline  
 RandHerd\cite{syta2017scalable}& A & \makecell{No ommission \\ in commit.}& \makecell{Prob.} & $O(c^2\log N)$ & $O(c^2\log N)$ & $f < \frac{N}{3}$ & \makecell{PVSS \cite{schoenmakers1999simple}\\ Multisignatures \cite{bellare2006multi}}\\ \hline 
 SCRAPE\cite{cryptoeprint:2017:216} & S & None & Det. & $O(N^3)$ & $O(N^2)$ & $f < \frac{N}{2}$ & PVSS \cite{schoenmakers1999simple}\\ \hline
 DFinity\cite{DBLP:journals/corr/abs-1805-04548}& S & None & \makecell{Prob.} & $O(cN)$ & $O(c)$ & $f < \frac{N}{2}$ & \makecell{BLS signatures \cite{boneh2001short}}\\ \hline 
 HydRand\cite{cryptoeprint:2018:319} & S & \makecell{No ommission \\ in commit.} & Det. & $O(N^2)$ & $O(N)$ & $f<\frac{N}{3}$ & PVSS \cite{schoenmakers1999simple}\\ \hline 
 ProofOfDelay\cite{bunz2017proofs} & S & None & Det. & \makecell{$O(N)$ + \\ Ethereum} & High & $f < \frac{N}{2}$ & Delay functions \cite{boneh2018verifiable}\\ \hline  
 No-Dealer\cite{krasnoselskii2020no}& S & None & Det. & $O(N^2)$ & $O(N^2)$ & $f<\frac{N}{2}$ & \makecell{Shamir \cite{shamir1979share}\\ Homomorphic Hash}\\ \hline 
 Nguyen et al.\cite{nguyen2019scalable} & S & \makecell{Trusted \\ Requester} & Prob. & $O(N)$ & $O(1)$ & $f < N$ & FHE \cite{gentry2009fully}, VRF \cite{micali1999verifiable}\\ \hline 
 \makecell{Ouroboros \\ Praos\cite{david2018ouroboros}} & P & \makecell{Weaker \\ properties} & Det. & \makecell{$O(N)$ + \\ Ourob. Praos} & \makecell{$O(1)$ + \\ Ourob. Praos} & $f<\frac{N}{3}$ & \makecell{VRF \cite{micali1999verifiable}}\\ \hline
 Algorand\cite{10.1145/3132747.3132757}& P & \makecell{Weaker \\ properties} & \makecell{Prob.} & \makecell{$O(cN)$ + \\ Algorand}& \makecell{$O(c)$ + \\ Algorand} & $f<\frac{N}{3}$ & \makecell{VRF \cite{micali1999verifiable}}\\ \hline
 \textbf{RandSolomon} & P & None & Det. & \makecell{$O(N) \times$ \\ Consensus} & \makecell{$O(N)\times $\emph{Erasure} \\ \emph{Correcting Code}} & $f < \frac{N}{3}$ & \makecell{PK crypto  \\ ReedSolomon \\ Retraceability}\\ [1ex] \hline
\end{tabular}}
\end{center}
    \caption{Comparison of distributed RNG solutions}
    \label{tab:comp}
\end{table}

\vspace{2mm}\noindent\emph{Synchrony (Sync).} The second column of the comparison table shows which kind of synchrony the underlying system must provide in order to allow the deployment of each protocol. Here we distinguish  A=Asynchronous, S=Synchronous and P=Partially Synchronous algorithms.

\vspace{2mm}\noindent\emph{Vulnerability.} It might seem impossible to have asynchronous implementations of \textsf{BFT-RNG} as we have already stated that this problem is impossible in the presence of at least one Byzantine participant in asynchronous systems~\cite{krasnoselskii2020no}. 
Notice, one might introduce additional assumptions on the failure model for these solutions to exist. 

This is the case with the solution by Cachin et al.~\cite{cryptoeprint:2000:034}  which assumes that there exists a special process capable of generating and distributing a key. 

Other asynchronous solutions, such as RandShare, RandHound and RandHerd~\cite{syta2017scalable}, assume that \emph{every} entity initially publishes some information about their secret. 
The asynchronous protocols in~\cite{syta2017scalable} are therefore not fully BFT, as they do not tolerate omission failures in the generation phase. 
This assumption that Byzantine processes will not omit during the commitment phase of the protocol is also an exploitable vulnerability in the \emph{synchronous} protocol HydRand~\cite{cryptoeprint:2018:319}, although it can be modified to restart once there are missing contributions.
Nguyen et al.'s proposal~\cite{nguyen2019scalable}, also a synchronous protocol, assumes a \emph{Requester}, a trusted entity generating FHE keys, which can be considered as a client using the system.

    Algorand~\cite{10.1145/3132747.3132757} and Ouroboros Praos~\cite{david2018ouroboros},  maintain weak forms of RNG: common coin~\cite{10.1145/3132747.3132757} and random beacon~\cite{david2018ouroboros},
%
%
RNG mechanisms in these protocols may not reach perfect agreement on the random value, and the coins values may be manipulated by the adversary to some extent or even be changed due to network asynchrony without affecting the correctness of their respective systems. 
%
%


\vspace{2mm}\noindent\emph{Termination (Term).}  A protocol ensures \emph{deterministic termination} (Det) if it  terminates in every execution,
in contrast to \emph{probabilistic termination} (Prob), when a protocol terminates with a fixed probability. 
 RandHound, RandHerd~\cite{syta2017scalable} and Dfinity~\cite{DBLP:journals/corr/abs-1805-04548} allow a small probability, depending on the parameters of the system, of the Byzantine adversary fully corrupting a cluster, which results in  prematurely halting the protocol. In the case of Algorand, a failure happens when the set (of expected cardinality $c$) of nodes chosen to be proposers is empty. 
In the protocol by Nguyen et al.~~\cite{nguyen2019scalable}, this happens when all selected contributors are Byzantine.

\vspace{2mm}\noindent\emph{Complexity.} \emph{Communication complexity} corresponds to the amount of messages exchanged and can be loosely translated into how many bits must be sent in the network for producing a result, while \emph{Computation Complexity} measures how much time would it take to perform local computations given an input. 
In the table, we use term \emph{High} to refer to  the complexity of delay functions, which, though independent of the number of processes in the system (strictly speaking, their complexity is $O(1)$), are very computationally heavy by design.

No-Dealer~\cite{krasnoselskii2020no} specifies that the protocol must be restarted in case of certain Byzantine  behavior, but does not include this fact in its complexity. 
As there are at most $\frac{N}{2}$ Byzantine nodes, it might be necessary to restart this number of times, increasing their claimed complexity to the one presented in the table.

The protocol by Nguyen et al.~~\cite{nguyen2019scalable}
employs a summation on the secrets shared by the contributors, which results in linear computation complexity.

Finally, the two last columns \textbf{Resilience} and \textbf{Techniques} show how many Byzantine processes can be tolerated among the $N$ participants and the main techniques employed in each solution.

\subparagraph*{Contributions.}
\textsf{RandSolomon} is the first \textsf{BFT-RNG} protocol providing deterministic termination in a partially synchronous system with $f<\frac{N}{3}$ Byzantine processes, which is the optimal level of  resilience~\cite{krasnoselskii2020no}.
Interestingly, the protocol relies only on standard cryptographic primitives: a public key infrastructure~\cite{salomaa2013public}, block erasure correcting codes which can be interpreted as our version of secret-sharing~\cite{mceliece1981sharing} and standard digital signatures. 
The name of the protocol is inspired by the potential use of Reed-Solomon codes~\cite{reed1960polynomial}.

Our coding approach carries some similarities with \textsf{SCRAPE}~\cite{cryptoeprint:2017:216} in the sense that they also recognised the potential of using codes such as Reed-Solomon to perform secret sharing.
However the similarities stop there as, in \textsf{RandSolomon}, we not only propose a partially synchronous solution, but also introduce a new technique to cope with Byzantine commission failures: \emph{retraceability}, which circumvents the need for verification of the secret sharing. 
In a nutshell, we consider the secrets produced by Byzantine processes without checking their integrity until the last phase of the protocol, when we compute the final result. 
At this moment, we can retrace all the steps that should have been taken and detect a commission failure. 
This then results in discarding incorrectly formed data in order to ensure a correct result, based on the inputs of non-Byzantine processes.

\section{Formal system model and properties}
\label{sec:prop}
Before turning to the \textsf{RandSolomon} protocol description, let us first duly formalise the system model as well as a set of properties that a protocol must have to be considered a distributed Byzantine fault-tolerant random number generator.

Our system is made up of $N$ nodes which run our protocol as a process which executes a prescribed sequence of steps. Among the participants, a portion $f < \frac{N}{3}$ of them might be Byzantine who can collaborate with each other but have limited computing power.

 The nodes can communicate with each other via messages that are sent through a point to point network. This network is available for all running processes and guarantees that if a message is sent through a channel, then it must be eventually delivered (in agreement with the partial-synchrony assumption). Whenever a process executes a broadcast it does so by just sending a message to every other process (we use a best effort broadcast).

\ignore{
Also, when a correct processes wishes to broadcast an information for the system, it does so by making means of a \emph{Reliable Broadcast} \cite{bracha1987asynchronous} primitive. Thanks to this, Byzantine processes cannot stop the spread of information in the system. We now build on this to define the following broadcast primitive.

\vspace{0.5cm}
\textbf{Byzantine Reliable Broadcast.}\\
\vspace{-0.5cm}
\begin{itemize}
    \item {\bf Validity} If $p_s$ is non-faulty and broadcasts $M$ in a byzantine reliably way then eventually some non-faulty process $p_i$ receives $M$.
    \item {\bf Integrity} If a non-faulty $p_i$ receives $M$ with correct sender $p_s$, then $p_s$ previously broadcast $M$ in a byzantine reliably way.
    \item {\bf Agreement} If a non-faulty process $p_i$ delivers $M$, then eventually every non-faulty process $p_j$ delivers $M$.
\end{itemize}
}


Recall that in a BFT-RNG protocol, every process proceeds through clearly demarcated phases: (1)~generation and commitment, (2)~reveal, and (3)~result computation.   
A phase begins with the first correct process entering it.
In this setup, a BFT-RNG protocol satisfies the following properties:

\begin{itemize}
    \item {\bf Agreement} Every correct process decides on the same random number;
    \item {\bf Unpredictability}  Before the beginning of the reveal phase, no process can distinguish an execution that generates $\mathit{RAND}$ as a random number,  from an execution that generates $\mathit{RAND'}$, for any $\mathit{RAND' \neq RAND}$; 
    \item {\bf Randomness}  The values decided by correct processes follow a uniform distribution; 
    \item {\bf Termination} Eventually, every correct process decides on a value.
\end{itemize}

 Although not an intrinsic property of \emph{BFT-RNG}s, our protocol differs from existing protocols because it provides \textbf{\textsc{retraceability}}. 
It means that after the reveal phase, a process can verify that all the steps taken to generate the shared data used to produce the final random number were correctly followed.

\section{The \textsf{RandSolomon} protocol}

\subparagraph*{Overview.} From a high-level viewpoint, the protocol aggregates enough locally generated random numbers, so that enough inputs inputs are truly random and the final result observes all the properties desired. 
Numbers are produced locally, then encoded using an erasure correcting code and encrypted before sharing. 
All non-Byzantine processes agree on which numbers should be used by solving consensus, while the result remains secret (sealed under an encryption layer) as no process holds all the information necessary for computing it prior to the reveal phase. 
The protocol cannot be stopped by $f$ (or less) Byzantine processes, as prior to the consensus the progress of correct processes depends solely on themselves and after it, thanks to our use of the erasure correcting code, the correct processes can retrieve data without using the information held by their Byzantine counterparts.

\subparagraph*{Notation.} We shall use $[N] = \{1,2,\cdots,N\}$, $(\cdot)_i$ to indicate that the value enclosed by the parenthesis contains a signature of process $p_i$ and $\{\cdot\}_i$ to indicate that the value enclosed by the curly brackets was encrypted using $p_i$'s public key. 
Furthermore, $b$ will denote the number of symbols in the encoded value to be encrypted in a given encryption key; $z$ the size of the symbols used in a code; $t$ is the number of erasures a code can correct; $l$ is the length of a code; $d$ the number of data symbols in a code.

\subsection{Primitives}
\label{sec:primitives}

The system requires a \textit{deterministic} encryption infrastructure where every process knows the pu\-blic key of every other processes in the system, but each of them maintains its private key  se\-cret. \textit{Deterministic} means here that at every time two processes encrypt the same number using the same key, they get the same result~\cite{bellare2007deterministic}.

Although the use of deterministic encryption is crucial for the correct execution of the protocol, these primitives are used only to encrypt long-enough (at least 256 bits) sequences of uniformly random bits. 
As such, the source of randomness in cleartext mitigates the security issues which crop up when using deterministic encryption~\cite{rackoff1991non}.


We use a \emph{consensus} protocol to ensure that each correct process disposes of the same information. 
The consensus protocol used here must ensure that  eventually every correct process outputs a value (Termination) and that not two correct processes outputs different values (Agreement).
Further, the protocol must ensure  
\emph{external validity}~\cite{cachin2001secure}: only a \emph{valid} value can be output,
i.e., the output must satisfy a predefined $\textit{valid}$ predicate: 
%
\begin{definition}[Predicate valid]
$\mathit{valid(v)}$ is true iff $v$ contains $N-f$ inputs signed by $N-f$ different processes.
\end{definition} 
Any partially-synchronous algorithm that tolerates $f$ Byzantine failures among $3f+1$ processes can be used~ \cite{buchman2018latest,yin2019hotstuff,10.1145/3132747.3132757}. 

Finally, let us consider a different perspective on secret sharing mechanisms \cite{shamir1979share}. 
In a classical Shamir secret-sharing protocol, when a dealer shares a secret $s$ with $N$ processes $p_1,p_2,\cdots, p_N$ using a threshold of $N-t$, it sends the shares $s_1, s_2, \cdots, s_N$ to their respective processes. 
Any $N-t$ of these shares are sufficient to retrieve $s$, while less than $N-t$ can reveal nothing on the secret in question.
Indeed, one could consider the string $s_1s_2\cdots s_N$ as a code, the non-received values as erasures and hence conclude that, in fact, the secret sharing scheme can be also analysed as an Erasure Correcting Code capable of correcting $t$ erasures \cite{mceliece1981sharing}.

In Information Theory, the number of substitutions required to change one string into another is known as \emph{Hamming Distance} \cite{macwilliams1977theory}. We can then conclude that we need in fact an Erasure Correcting Code with Hamming distance at least $t+1$. The class of error-erasure correcting codes known as \emph{Reed-Solomon (RS)}\cite{reed1960polynomial} with the required distance is capable of correcting $t$ \textbf{erasures} (notice we do not treat it as an error correcting code, but an erasure correcting: an error correcting code is capable of correcting a string with corrupted data placed in unknown locations, while an erasure correcting code needs to know the positions of the string which were corrupted). 
Therefore, this class provides optimal block size known as \emph{Singleton Bound} \cite{singleton1964maximum}. 
From a more pragmatic viewpoint, Reed-Solomon codes have free library implementations in many programming languages, they have deterministic parameters and encoding which are ideal for our requirements. 
Furthermore, most applications running our protocol will have relatively small block sizes and one can enhance the performances through hardware implementations \cite{khan2003hardware}. It should be noted however, that any code complying with the following \emph{Abstract Code} requirements can be used in our protocol. 

\vspace{0.25cm}
\textbf{Abstract Code.}\\
\vspace{-0.5cm}
\begin{itemize}
    \item Have a code-word of size $b\times N$ symbols;
    \item Be able to correct up to $b \times f$ symbol erasures;
    \item $b \times z \geq 256$
\end{itemize}
\label{Desired ECC}

Considering that we make use of Reed-Solomon codes we briefly present their general parameters:

\vspace{0.25cm}
\textbf{Abstract Reed-Solomon code}\\
\vspace{-0.5cm}
\begin{itemize}
    \item The symbols have size $z$ bits
    \item The data has length $d$ symbols
    \item The code-word has length $l$ where $l \leq 2^z-1$ symbols
    \item It can correct up to $t$ erasures where, $t=l-d$
\end{itemize}

Adjusting the above \emph{Abstract RS code} to match the \emph{Abstract code} and the system requirements, leads to the following Concrete Reed-Solomon Code which is suitable for implementing our protocol. 

\vspace{0.25cm}
\textbf{Concrete Reed-Solomon code}\\
\vspace{-0.5cm}
\begin{itemize}
    \item The symbols have size $z$ bits;
    \item Each block to be encrypted has a size $b$ of at least $\frac{256}{z}$ symbols;
    \item The data has a length of $b(N-f)$-symbols;
    \item The code-word has a length of $b \times N$ symbols.
\end{itemize}

It should be noted that as our protocol allows correct processes to retrace the execution followed by Byzantine processes and detect when they generate incorrect messages, we can use erasure correction instead of error correction. 
This drastically improves the coding performance as every error-erasure correcting code can correct two times more erasures than errors. 
This has two implications on our protocol: first we need fewer parity bits; second, if we were to unnecessarily use the code for errors correction, the protocol would only tolerate up to $\lfloor \frac{N-1}{4} \rfloor$ Byzantine processes. 
The reason for the potential loss of resilience comes from the fact that we would need to correct $2f$ errors: $f$ errors introduced by the Byzantine member during the generation and $f$ more for the missing blocks due to asynchrony. 
Therefore the number of parity blocks would have to be at least $2(2f)=4f$ blocks, while the code must have length $N$ blocks. Because the length of a code is larger than the number of parity symbols, $N > 4f$. This illustrates the contribution of retraceability: it implies simpler data reception by eliminating the need to generate proofs and to check them, and guarantees better resilience whilst maintaining the correctness of the protocol.

\subsection{Algorithm}
\label{sec:alg}
\begin{figure}[!htb]
\hrule \vspace{1mm}
 {
\begin{tabbing}
bbbb\=bbbb\=bbbb\=bbbb\=bbbb\=bbbb\=bbbb\=bbbb \=  \kill
\textit{Each function is entirely executed before executing the next}\\
\textbf{Static Local Variables:}\\
\textit{RNL} := 0: set of encoded and encrypted shared random numbers learnt in Consensus\\
\textit{SEEN} := $\emptyset$: map where the key is the index of a process and the value is the value it produced\\
$\mathit{\sigma_i[1..N][1..N]} := \bot$: array of plain random number shares used in reconstruction\\
$\mathit{RAND_i} := 0$: random number decided by $p_i$\\
\\
\textbf{\{Generation and Commitment Phase\}}\\
\nnll\label{line:rs:localrand}\> Generate random number $r_i$ of $b(N-f)$ symbols of $z$-bits\\
\nnll\label{line:rs:rsencrand}\> Encode $r_i$ into $s_i$ with Desired RS\\
\nnll\label{line:rs:pkencrand}\> $\underline{s_i} = (\{s_i[1]\}_1,\{s_i[2]\}_2, \cdots, \{s_i[N]\}_N)_i$\\
\nnll\label{line:rs:bcastrand}\> Broadcast $\langle \mathit{GENERATED}, \underline{s_i} \rangle$\\
\\
\textbf{upon} receiving $\langle \mathit{GENERATED}, \underline{s_j} \rangle$\\
\nnll\label{line:rs:registervalue}\> $SEEN[j]:=\underline{s_j}$\\
\nnll\label{line:rs:proposerand}\> \textbf{if} $\vert SEEN \vert = N-f$ \textbf{then} $\mathit{RNL} :=$ $\mathit{Consensus}(SEEN)$\\
\\
\textbf{\{Reveal Phase\}}\\
\textbf{upon} $RNL \neq \emptyset$\\
\nnll\label{line:rs:loopRNL}\> $\forall \underline{s_j} \in RNL$ \textbf{do}\\
\nnll\label{line:rs:decShare}\>\> Decrypt $\underline{s_j}[i]$ from $\underline{s_j}$ into $s_j[i]$\\
\nnll\label{line:rs:assignDec}\>\> $\sigma_i[j][i] := s_j[i]$\\
\nnll\label{line:rs:reveal}\> Broadcast $\langle \mathit{REVEAL}, (\sigma_i[:][i])_i \rangle$\\
\\
\textbf{upon} receiving $\langle \mathit{REVEAL}, (\sigma_j)_j \rangle$, $j \neq i$ \textbf{execute after} $RNL \neq \emptyset$\\
\nnll\label{line:rs:loopReveal}\> $\forall \underline{s_k} \in RNL$ \textbf{do}\\
\nnll\label{line:rs:checkDec}\>\> \textbf{If} $\{\sigma_j[k][j]\}_j = \underline{s_k}[j]$ from $\underline{s_k}$ \textbf{then} $\sigma_i[k][j] := \sigma_j[k][j]$\\
\\
\textbf{\{Result Computation Phase\}}\\
\textbf{upon} $RNL \neq \emptyset \wedge \forall \underline{s_j} \in RNL, \exists K \subseteq [N], \vert K \vert = N-f: \sigma_i[j][k] \neq \bot$\\
\nnll\label{line:rs:}\> $\mathit{step}:=0$\\
\nnll\label{line:rs:preRand}\>\> $PRE := 0$\\
\nnll\label{line:rs:loopReconstruct}\> $\forall \underline{s_j} \in RNL$ sorted by $j$ \textbf{do}\\
\nnll\label{line:rs:decodeRS}\>\> Decode $\sigma_i[j]$ into $\tilde r_j$ using Desired RS\\
\nnll\label{line:rs:checkIntegrity}\>\> \textbf{If} $\tilde r_j$ encoded with Desired RS and blockwise encrypted doesn't match $\underline{s_j}$ \textbf{then} $\tilde r_j := 0$\\
\nnll\label{line:rs:produceRand}\>\> $PRE :=  PRE \oplus (\tilde r_j \gg step)$ \{Circular right shift by \textit{step} blocks or $b\times \mathit{step}$ symbols\}\\
\nnll\label{line:rs:incStep}\>\> $step{+}{+}$\\
\{XOR blocks pairwise with triple in the end if necessary\}\\
\nnll\label{line:rs:pairwise}\> \textbf{for} $k := 1;~2k-1 < N-f;~k:=k+2$\\
\nnll\label{line:rs:rand}\>\> $RAND_i[k] := PRE[2k-1] \oplus PRE[2k]$\\
\nnll\label{line:rs:triple}\> \textbf{if} $2k-1 = N-f$ \textbf{then} $RAND_i[k] := RAND_i[k] \oplus PRE[N-f]$\\
\nnll\label{line:rs:decideRand}\> Decide $RAND_i$
\end{tabbing}
}
\vspace{-1.5mm}
 \hrule
\caption{RandSolomon code for process $p_i$}
\label{fig:RandSolomon}
\end{figure}

\subparagraph*{Generation and commitment.} Each process $p_i$ taking part in the protocol begins by generating a random number $r_i$ of $b(N-f)$ symbols and encoding it using a Reed-Solomon encoder complying with the specification given in \autoref{sec:primitives} obtaining a number $s_i$ of $b\times N$ symbols (lines \ref{line:rs:localrand}, \ref{line:rs:rsencrand}). This encoded number $s_i$ is then split in $N$ blocks of $b$ symbols and each of these blocks are encrypted using the public key of the different processes in the system in order, signing the final result and obtaining the variable $\underline{s_i}$ (line \ref{line:rs:pkencrand}). 

Each process $p_i$ share their $\underline{s_i}$ (line \ref{line:rs:bcastrand}) and collect $N-f$ numbers of this type, coming from $N-f$ distinct processes according to their signatures. With this set of $N-f$-numbers they can engage in consensus and learn the same set, say \emph{RNL}, of $(N-f)$ numbers generated by $N-f$ distinct processes (line \ref{line:rs:proposerand}).

\subparagraph*{Reveal.} After obtaining the \emph{RNL} set, each process can decrypt the blocks it is responsible for (line \ref{line:rs:decShare}) and reveal them to the system via a broadcast (line \ref{line:rs:reveal}). 

(A best-effort broadcast in which a process simply sends the message to every other process will suffice.)  

The processes gather the shares necessary for decoding the erasure correcting code, making sure that they truly are the decrypted versions of the RNL shares (line \ref{line:rs:checkDec}).

\subparagraph*{Result computation.} Once a process has gathered at least $N-f$ shares of each of the numbers in the RNL set, it can reconstruct all of them (line \ref{line:rs:decodeRS}). If the decoded version $\tilde r_j$ of a RNL number is again encoded and encrypted, leading to the same value for $\underline{s_j}$, then this implies that any $N-f$ shares obtained by any correct process will give the same $\tilde r_j$ making it consistent to be used in the final step computations.

\subparagraph*{Importance of verification.} Notice that if $p_i$ is Byzantine, then it can generate a number $r_i$ and insert $f$ blocks with errors in $s_i$.  By colluding with other Byzantine processes in the system, a correct process $p_j$ might get no response from $f$ Byzantines and get these $f$ erroneous blocks, essentially receiving a number with $2f$ incorrect blocks, which leads it to decode a number $\tilde r_i' \neq r_i$. Meanwhile a process $p_k$ can get the Byzantine processes' correct shares instead of the blocks with errors, decoding $\tilde r_i'' = r_i$, which would lead these two different correct processes producing two different random numbers in the end. This attack is nullified by the simple verification done in the line \ref{line:rs:checkIntegrity}
and setting this number produced by a Byzantine process to 0, which is done by every process. It should be noted that because at least $N-f$ numbers are used and that there are at most $f$ Byzantines, at least $f+1$ numbers will not be nullified. 

 \subparagraph*{Cyclic XOR.} Finally the correct processes will hold the same decoded versions of the RNL numbers which are well formed and can produce the same final random number by first cyclically shifting each number to the right by increasing steps of blocks (remember a block has $b$ symbols) and then taking an XOR of them (line \ref{line:rs:produceRand}). Here, the reason for the shift is that for Byzantine processes might know the full contents of up to $f$ numbers and $f$ positions from each of the other numbers before the reveal phase. Assuming all the numbers produced by Byzantines were chosen, then the shift ensures that at least $f+1$ different positions from the numbers created by correct processes will be used, hence including at least one unknown value for the malicious participant before the reveal.

\subparagraph*{Pairwise (triple) XOR and decision.} The final step is to XOR the last three blocks together and the remaining blocks pairwise when $N-f$ is odd and XOR all the blocks pairwise when $N-f$ is even. Suppose this last step was not taken and the shifted XOR blocks were returned. Then if the $2f$ positions known by the Byzantine could potentially be used in the computation of a position $pos$ in the result and these blocks XOR to a value $x$, they can assure that by promoting any unknown value different than $x \oplus y$ to be the last operand used in $pos$ assures that the value $y$ will not appear in $pos$. Because of the deterministic encryption they can immediately check the candidate values for being different than $x \oplus y$, although it is computationally unfeasible to determine their value. In our solution, however, because we guarantee that the Byzantine do not know at least two values used, there are $2^{b\times z}$ pairs that XOR to any given value and it is unfeasible to test the two values for being different than all of them (as $b\times z \geq 256$ in real scale instantiations of the protocol), let alone read, which would take $2^{2\times b \times z}$ tests.

\subsection{Execution example}

\begin{figure}

\begin{subfigure}[b]{0.8\textwidth}
\centering
   \caption{}
   \includegraphics[width=0.8\textwidth]{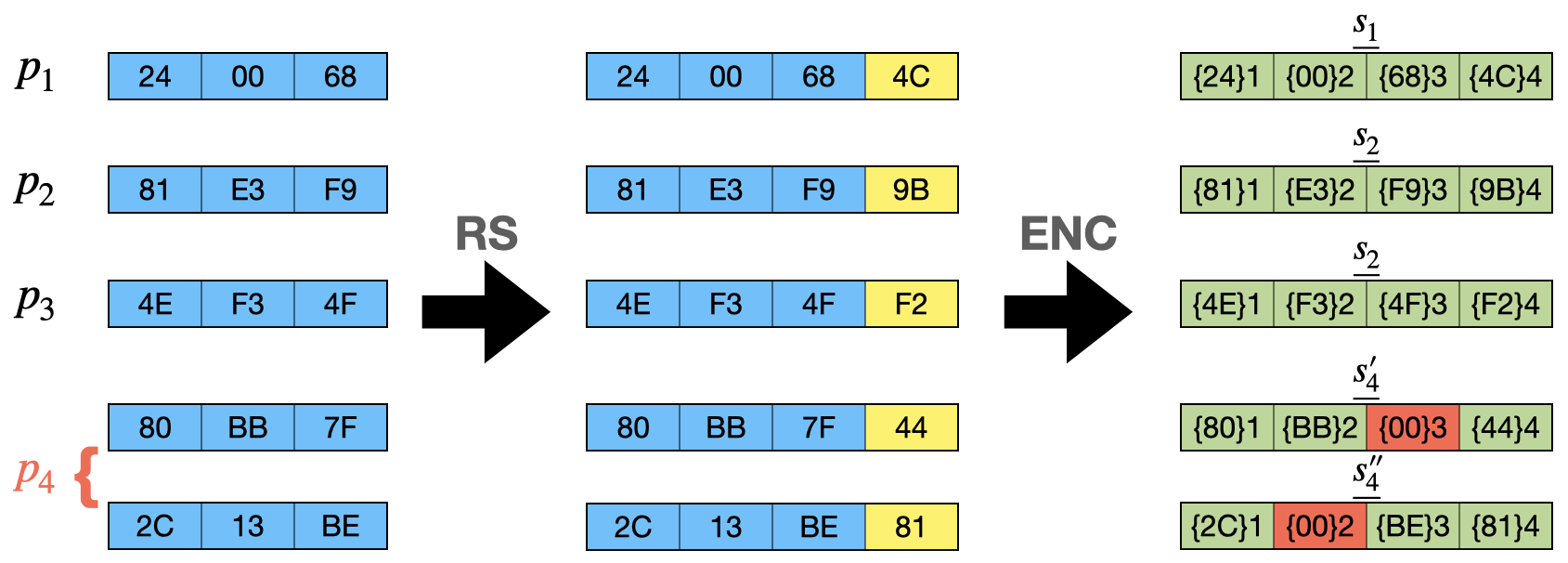}
   \label{fig:rs1} 
\end{subfigure}

\begin{subfigure}[b]{0.8\textwidth}
\centering
   \caption{}
   \includegraphics[height=0.15\textheight]{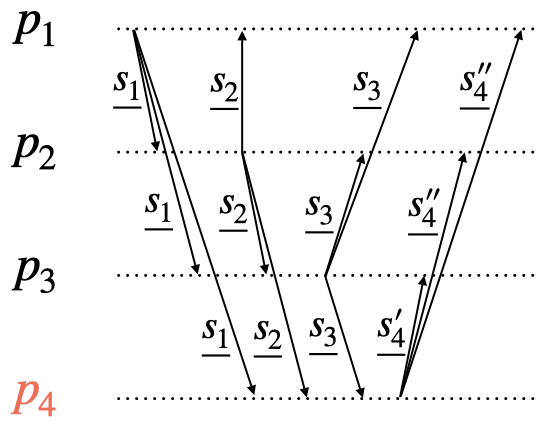}
   \label{fig:rs2} 
\end{subfigure}

\begin{subfigure}[b]{0.8\textwidth}
\centering
    \caption{}
    \includegraphics[width=0.85\textwidth]{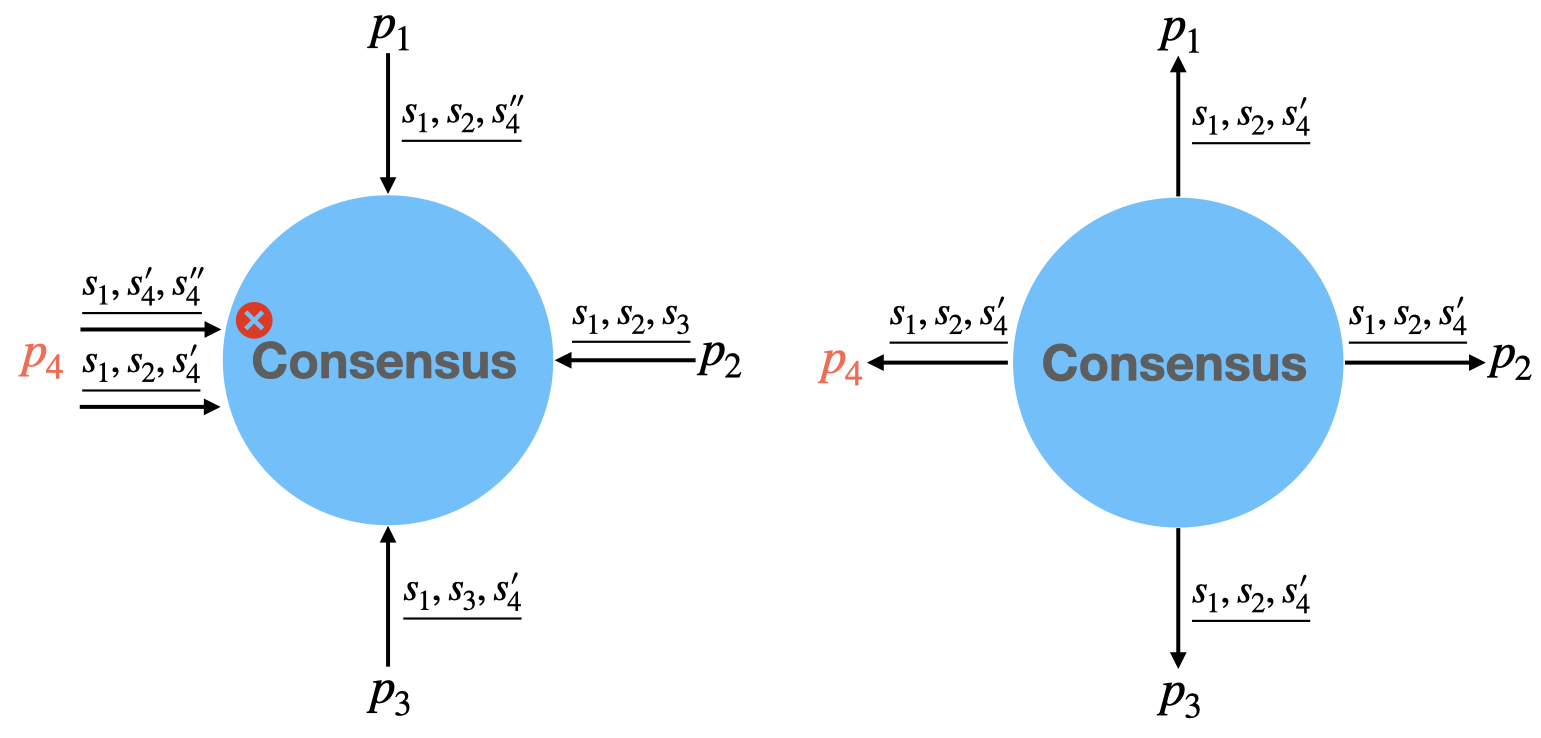}
   \label{fig:rs3} 
\end{subfigure}

\begin{subfigure}[b]{0.8\textwidth}
\centering
    \caption{}
    \includegraphics[height=0.15\textheight]{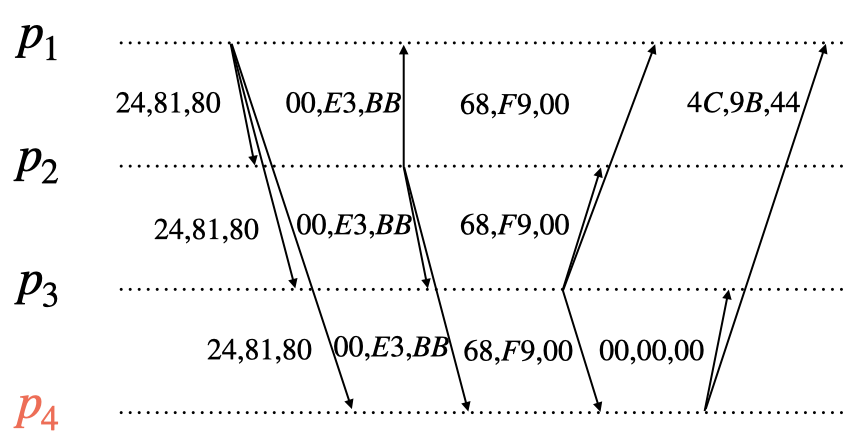}
   \label{fig:rs4} 
\end{subfigure}

\begin{subfigure}[b]{0.8\textwidth}
\centering
   \caption{}
   \includegraphics[width=0.8\textwidth]{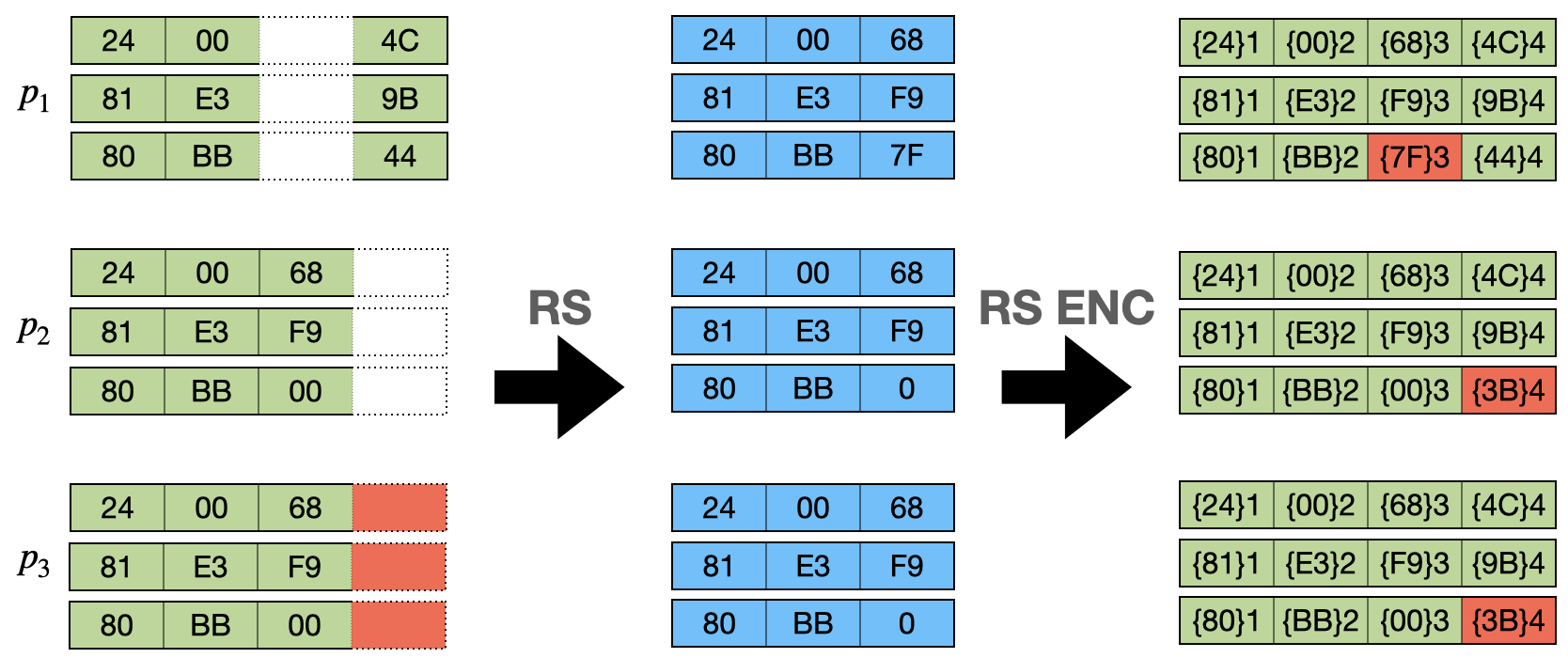}
   \label{fig:rs5} 
\end{subfigure}
\end{figure}

\begin{figure}\ContinuedFloat

\begin{subfigure}[b]{0.8\textwidth}
\centering
   \caption{}
   \includegraphics[width=0.7\textwidth]{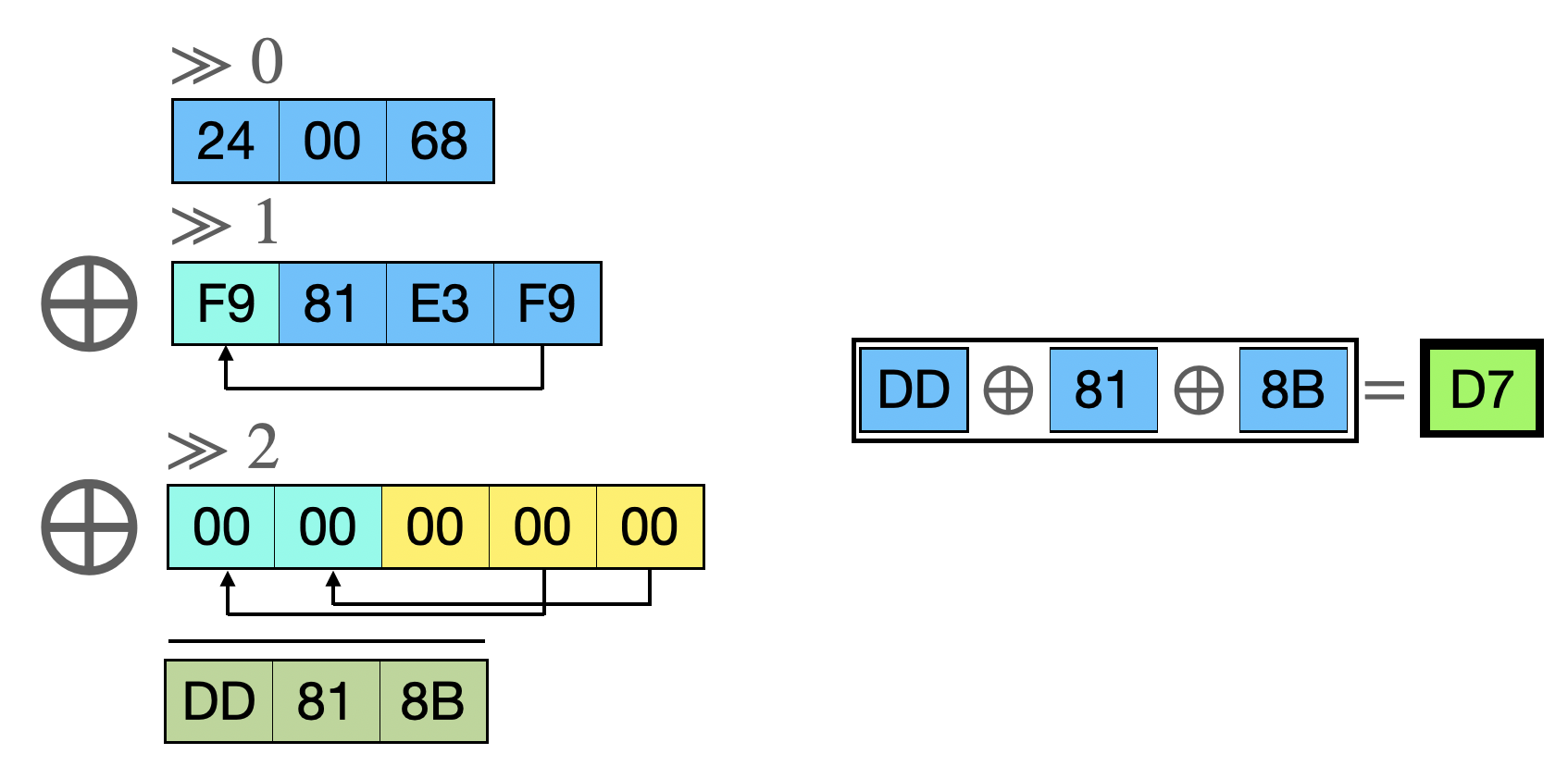}
   \label{fig:rs6} 
\end{subfigure}

\caption{Example of a RandSolomon execution with one Byzantine process among a system of 4 processes.}
\label{fig:eg}
\end{figure}

We present now an example of a possible execution of our protocol with one Byzantine process and four processes in total illustrated in \autoref{fig:eg}. For pedagogical reasons we assume that the symbols have $8$-bits and that each block to be encrypted contains $1$ symbol ($b=1,z=8$), relaxing the requirement that $b\times z \geq 256$.

The beginning of the protocol and the \emph{Generation and Commitment Phase}, corresponding to lines from lines \ref{line:rs:localrand} to \ref{line:rs:pkencrand} of the algorithm is shown in \autoref{fig:rs1}. The correct processes $p_1,p_2$ and $p_3$ produce each a $3$ bytes random number, correctly encoding into a $4$ bytes reed-solomon codeword. The values $\underline{s_1},\underline{s_2},\underline{s_3}$ ready to be shared are obtained by encrypting each of the 4 bytes from the codewords with the public keys of the $p_1,p_2,p_3$ and $p_4$, respectively. On the other hand, process $p_4$, who is Byzantine, maliciously produces two bad values: $\underline{s'_4}$ with an error in its third byte and $\underline{s''_4}$ with an error on its second byte.

\autoref{fig:rs2} then shows lines \ref{line:rs:bcastrand} and \ref{line:rs:registervalue} where processes share their produced values and collect values coming from other processes. Notice that contrary to correct processes, Byzantine processes might send different values to different destinations.

 Once each process has gathered three (N-f) different values, they propose what they know to the consensus component (line \ref{line:rs:proposerand} and \autoref{fig:rs3}). Nothing prevents the Byzantine process $p_4$ of making more than one proposal to consensus, but any proposal which is not composed by $N-f$ signatures is discarded. Once the consensus algorithm terminates, any valid value might be returned, but all processes will get the same result (decided value equal to $\underline{s_1}, \underline{s_2}$ and $\underline{s'_4}$). 

The \emph{Reveal Phase} illustrated in \autoref{fig:rs4} then begins, comprising lines \ref{line:rs:loopRNL} to \ref{line:rs:checkDec}. At this point processes openly share the symbols that were previously encrypted in their public keys. One deviation Byzantine processes might do is to send wrong numbers that do not correspond to the agreed values counterparts. However, because of the deterministic encryption, the receiver can detect it by asserting that the encrypted version does not match the plain value received and discard it. Moreover, even if the Byzantine process does not send its share to every participant it does not matter, as $N-f$ shares are available nonetheless. 

Once processes gather three shares for each of the numbers agreed upon in consensus they can start the \emph{Result Computation Phase} executing lines \ref{line:rs:loopReconstruct} to \ref{line:rs:decideRand}. \autoref{fig:rs5} shows how they first obtain the decoded version of the numbers and then redo both the reed-solomon encoding and the encryption of the blocks to check that they correspond to the value decided in consensus. At this point they discard the value generated by $p_4$ nullifying its contribution and computing the final random number by XORing the other values as shown in \autoref{fig:rs6}.

 On the right column of the same figure we can see that the processes sort the agreed numbers by their origin, in this case they take $r_1,r_2$ and $r_4''$ in this order. They proceed by cyclically shifting the first number by $0$ blocks, the second by $1$ block and the last by $2$ blocks. They obtain the same number $DD,81,8B$ and produce the same final random number $D7$ by XORing all three blocks, as these are the last three blocks. Note that if the system had $f=3$ and $N=10$, for example, the result from the cyclic XOR would have $N-f=7$ blocks $B_1 | B_2 | B_3 | B_4 | B_5 | B_6 | B_7$ and the final random number would have three blocks: $B_1 \oplus B_2 | B_3 \oplus B_4 | B_5 \oplus B_6 \oplus B_7$. 

\section{Formal analysis of the protocol}
\subsection{Correctness}
This section is devoted to the proof that \textsf{RandSolomon} is a correct partially-synchronous BFT-RNG. We do so by showing that the protocol satisfies the set of properties stated in Section \ref{sec:prop}.

\setcounter{theorem}{0}

\begin{proposition}
\textsf{RandSolomon} achieves \emph{Agreement}.
\begin{proof}
 Because of the consensus using the external validity property , every correct process has the same \emph{RNL} set. Correct processes then use shares that have been verified and match the values agreed upon (line \ref{line:rs:checkDec}), allowing them to only access the original values generated in line \ref{line:rs:rsencrand}.

If a RNL number $\underline{s_j}$ passes the test in line \ref{line:rs:checkIntegrity}, any $N-f$ correctly decrypted shares of this number shall yield the same number, as the encoded value contains no errors. It follows that every correct process will only use correctly decrypted shares and every correct process will hold the same number $\tilde r_j$ which will pass the test by our hypothesis.

If, however, this RNL number $\underline{s_j}$ does not pass the test, then there is an error in its encoding, as the test is merely checking if it was correctly done, and it will be visible to all correct processes in the system which will all proceed to ignore this number.

 Therefore all $\mathit{RAND_i}$ are equal, as they are formed by XORing and shifting the same RNL numbers which every correct process agrees upon. 

\end{proof}
\end{proposition}

\begin{proposition}
\textsf{RandSolomon} achieves \emph{Unpredictability}.
\begin{proof}

A process with limited computational power has negligible probability of determining the plain value corresponding to an encrypted value it does not possess the decryption key of. It can however test that it does not correspond to a certain value.

If Byzantine processes collude and share each others values before the different processes agree on which $N-f$ values at the end of the generation phase will compose the final result, they will know at most $f$ full values. They will also possess $f$ shares of each of the remaining $f+1$ chosen values corresponding to their positions but it is impossible for them to get any more shares prior to correct processes entering the reveal phase and sending them this information. Thus, they cannot determine the value of any given position in the decided value as the shifts makes so that at least $2f+1$ positions from the operands are needed in order to determine a position from the result and as established, the Byzantine can know at most $2f$ of them. It can still determine that the result is different than some specific value though, but as each position is then determined by the XORed with at least one other position, this possibility is then nullified as it would require the Byzantine processes to test $2^{b\times z}$ pairs of numbers in order to eliminate a value, which is computationally unfeasible with real scale protocol parameters ($b\times z \geq 256$).

\end{proof}
\end{proposition}

\begin{proposition}
\textsf{RandSolomon} achieves \emph{Randomness}.
\begin{proof}

By hypothesis, correct processes are capable of generating uniformly random numbers. The result of XORing a uniformly distributed random variable $X$ in $D$ with a constant $c$ in $D$ is a uniformly distributed random variable in $D$. Also, the result of XORing two independent uniformly distributed variables $X$ and $Y$ over $D$ is uniformly distributed. As we already established in the final two paragraphs of \autoref{sec:alg}, each position in the final result is independent from each other and uses at least two uniform random numbers coming from correct processes unknown to the Byzantine before the reveal phase. This means no proposed values are preferred over others and the randomness of the operands is transferred to the output.

\end{proof}
\end{proposition}

\begin{proposition}
\textsf{RandSolomon} achieves \emph{Termination}.
\begin{proof}

Every correct process generates their random numbers and propose a set of $N-f$ of them to the consensus component. This means that there will be at least $N-f$ processes engaging in it, and because it can tolerate up to $f$ failures, it will eventually give all correct processes their RNL sets.

Once $N-f$ correct processes learn what the RNL set is, they will share their shards, meaning that each correct process is guaranteed to receive at least $N-f$ correct shares of each of their RNL numbers, satisfying the conditions for entering the computation phase, where their progress becomes purely local as they do not depend on other processes anymore.
\end{proof}
\end{proposition}

\subsection{Complexity}
We shall analyse our algorithm in terms of \emph{message complexity}: the maximum number of messages transmitted per random number generated; \emph{bit complexity}: the maximum number of bits exchanged over the network per random bit generated; \emph{time complexity}: the number of message round trips required per random number generated; and \emph{computational complexity}: the number of operations to be executed per process per random number generated.

In the generation and commitment phase, each process executes one broadcast, meaning that there are $O(N^2)$ messages being sent at this phase. After consensus is reached on the value of \emph{RNL}, each process executes exactly one more broadcast, leaving the message complexity of this part of the protocol on $O(N^2)$. The result computation phase in done locally. Hence the message complexity of our protocol is $O(N^2)$ outside consensus.

In terms of bit complexity, \textsf{RandSolomon} produces random numbers of $O(N)$ bits, therefore we consider the number of bits exchanged divided by $N$. The messages of the generation phase contain random numbers whose lengths are proportional to the number of processes in the system by design. Therefore, the bit complexity of this step is $O(N^2)$. Afterwards in the reveal phase, each process includes one decrypted block per number in the \emph{RNL} set. Each decrypted block has constant size and the cardinality of \emph{RNL} is $f+1$, so the bit complexity of this stage is also $O(N^2)$. Therefore, without taking consensus into account, the bit complexity of our protocol is $O(N^2)$. The inputs for consensus are comprised of $N-f$ values of $O(N)$ bits and therefore the bit complexity (used in the Table \ref{tab:comp}) is $O(N)\times$\emph{Consensus}.

With respect to time complexity, our protocol requires outside consensus two message delays given the two aforementioned broadcasts, each executed by all processes in parallel. Consensus might require \emph{view-changes} in the worst case bringing its time complexity to $O(f)$ which corresponds to our overall time complexity. As for computational complexity we present the analysis split on the three phases of the protocol in \autoref{tab:complexity}. 

\begin{table}[h]
    \begin{center}
 \begin{tabular}{c | c | c | c}
 \hline
  Operation &  Generation & Reveal & Result \\
  \hline \hline
  Encryption & $O(N)$ & $O(N^2)$ & $O(N^2)$ \\ \hline
  Decryption & 0 & $O(N)$ & 0 \\ \hline
  ECC encoding & $O(1)$ & 0 & $O(N)$ \\ \hline
  ECC decoding & 0 & 0 & $O(N)$ \\ \hline
\end{tabular}
\end{center}
    \caption{Computational complexity}
    \label{tab:complexity}
\end{table}

If the erasure correcting code used is indeed Reed-Solomon, then the encoding and decoding complexities of a single number with length $O(N)$ is $O(N\log N)$ \cite{soro2010fnt}, meaning that the per-process computational complexity is $O(N^2\log N)$ when this particular code is used.

When considering the complexity of the Consensus protocol, one can easily adopt last generation PBFT consensus protocols developed in the context of blockchain-type ledgers. 
In this context, the Tendermint (analysed in detail in \cite{amoussou2019dissecting}) or Hotstuff \cite{yin2019hotstuff} consensus protocols can be used within \textsf{RandSolomon}. 
Doing so leads to an overall message complexity of $O(N^2)$ and bit complexity of $O(N^3)$ accounting view-changes with the complexity of consensus dominating that of our protocol for both protocols considered. 
As such, any system which already has the protocol machinery to solve consensus can implement \textsf{RandSolomon} without incurring a significant performance impact. 

In a run where the system is synchronous (after passed GST) and the consensus leader is correct, the protocol terminates in constant number of message delays and incurs only $O(N^2)$ bit complexity, comparable to that of synchronous protocols.

The complexity analysis of RandSolomon is summarised in \autoref{tab:fullcomplexity}.

\begin{table}[ht]
    \begin{center}
 \begin{tabular}{c | c | c | c | c | c}
 \hline
  Complexity & Generation & Consensus & Reveal & Result & Total\\
  \hline \hline
  Message & $O(N^2)$ & $O(N^2)$ & $O(N^2)$ & $0$ & $O(N^2)$\\ \hline
  Bit & $O(N^2)$ & $O(N^3)$ & $O(N^2)$ & 0 & $O(N^3)$ \\ \hline
  Time & 1 msg delay & $O(f)$ & 1 msg delay & 0 & $O(f)$ \\ \hline
  Computation & $O(N)$ & $O(N)$ & $O(N^2)$ & $O(N^2\log N)$ & $O(N^2\log N)$\\ \hline
\end{tabular}
\end{center}
    \caption{RandSolomon Protocol complexities integrating Consensus as in \cite{yin2019hotstuff}}
    \label{tab:fullcomplexity}
\end{table}

\section{Conclusion}



We presented \textsf{RandSolomon}, a Byzantine fault-tolerant protocol capable of generating a common random number in a partially-synchronous system. 
As we have previously shown in \autoref{sec:related}, although the problem of generating randomness in multi-party systems has already been extensively discussed, the partially-synchronous systems still lacked a BFT solution with the optimal resilience of $f$ Byzantine participants among $3f+1$ with deterministic termination. 
Not only did we provide such a solution but we also employed very simple public key cryptography, not relying on a random oracle, by means of what we have called \emph{retraceability}. Our approach is modular, using Consensus as a black box, which facilitates future implementations of the protocol with improved complexity metrics.


\bibliographystyle{abbrv}
\bibliography{references}

\end{document}